\documentclass[aps,prl,reprint,superscriptaddress]{revtex4-1}
\usepackage{todonotes}
\usepackage{graphicx,xcolor}
\usepackage{amsmath}
\usepackage{braket}
\usepackage{url}
\usepackage{filecontents}

\usepackage[pdftex,colorlinks=true,hypertexnames=false]{hyperref}

\newcommand{\NHO}{Nd$_2$Hf$_2$O$_7$ }
\newcommand{\AIAO}{all-in--all-out }

\newcommand{\vect}[1]{\mathbf{#1}}

\def\sectionn#1{\noindent\underline{\it #1:}}

\begin{document}


\title{Inverted hysteresis and negative remanence in a homogeneous
  antiferromagnet}


\author{L. Opherden} \email[]{l.opherden@hzdr.de}
\affiliation{Hochfeld-Magnetlabor Dresden (HLD-EMFL), Helmholtz-Zentrum
  Dresden-Rossendorf, 01328 Dresden, Germany} \affiliation{Institut f\"{u}r
  Festk\"{o}rper- und Materialphysik, TU Dresden, 01062 Dresden, Germany}
\author{T. Bilitewski} \affiliation{Max-Planck-Institut f\"{u}r Physik komplexer
  Systeme, 01187 Dresden, Germany} \author{J. Hornung}
\affiliation{Hochfeld-Magnetlabor Dresden (HLD-EMFL), Helmholtz-Zentrum
  Dresden-Rossendorf, 01328 Dresden, Germany} \affiliation{Institut f\"{u}r
  Festk\"{o}rper- und Materialphysik, TU Dresden, 01062 Dresden, Germany}
\author{T. Herrmannsd\"orfer} \affiliation{Hochfeld-Magnetlabor Dresden
  (HLD-EMFL), Helmholtz-Zentrum Dresden-Rossendorf, 01328 Dresden, Germany}
\author{A. Samartzis} \affiliation{\mbox{Helmholtz-Zentrum Berlin f\"{u}r
    Materialien und Energie GmbH, Hahn-Meitner Platz 1, 14109 Berlin, Germany}}
\affiliation{\mbox{Institut f\"{u}r Festk\"{o}rperphysik, Technische
    Universit\"{a}t Berlin, Hardenbergstra$\beta$e 36, 10623 Berlin, Germany}}
\author{A.~T.~M.~N. Islam} \affiliation{\mbox{Helmholtz-Zentrum Berlin f\"{u}r
    Materialien und Energie GmbH, Hahn-Meitner Platz 1, 14109 Berlin, Germany}}
\author{V.~K. Anand} \affiliation{\mbox{Helmholtz-Zentrum Berlin f\"{u}r
    Materialien und Energie GmbH, Hahn-Meitner Platz 1, 14109 Berlin, Germany}}
\author{B. Lake} \affiliation{\mbox{Helmholtz-Zentrum Berlin f\"{u}r Materialien
    und Energie GmbH, Hahn-Meitner Platz 1, 14109 Berlin, Germany}}
\affiliation{\mbox{Institut f\"{u}r Festk\"{o}rperphysik, Technische
    Universit\"{a}t Berlin, Hardenbergstra$\beta$e 36, 10623 Berlin, Germany}}
\author{R. Moessner} \affiliation{Max-Planck-Institut f\"{u}r Physik komplexer
  Systeme, 01187 Dresden, Germany} \author{J. Wosnitza}
\affiliation{Hochfeld-Magnetlabor Dresden (HLD-EMFL), Helmholtz-Zentrum
  Dresden-Rossendorf, 01328 Dresden, Germany} \affiliation{Institut f\"{u}r
  Festk\"{o}rper- und Materialphysik, TU Dresden, 01062 Dresden, Germany}

\date{\today}

\begin{abstract}%
  Magnetic remanence -- found in bar magnets or magnetic storage devices -- is
  probably the oldest and most ubiquitous phenomenon underpinning technological
  applications of magnetism. It is a macroscopic non-equilibrium phenomenon: a
  remanent magnetisation appears when a magnetic field is applied to an
  initially unmagnetised ferromagnet, and then taken away. Here, we present an
  inverted magnetic hysteresis loop in the pyrochlore compound
  Nd$_2$Hf$_2$O$_7$: the remanent magnetisation points in a direction {\it
    opposite} to the applied field. This phenomenon is exquisitely tunable as a
  function of the protocol in field and temperature, and it is reproducible as
  in a quasi-equilibrium setting. We account for this phenomenon in considerable
  detail in terms of the properties of non-equilibrium population of domain
  walls which exhibit a magnetic moment between domains of an ordered
  antiferromagnetic state which itself has zero net magnetisation. Properties
  and (non-equilibrium) dynamics of topological defects play an important role
  in modern spintronics, and our study adds an instance where a uniform field
  couples selectively to domain walls rather than the bulk.
\end{abstract}


\maketitle

The 'all-in--all-out' state of pyrochlore magnets plays an important role in the
study of topological magnets on account of its role in the genesis of
condensed-matter axion electrodynamics \cite{Wan_2011}. Here, we study its
response to an applied field in the insulating magnetic pyrochlore compound
Nd$_2$Hf$_2$O$_7$.

We report the observation of a fully inverted magnetic hysteresis loop probed
by the dynamic susceptibility resulting in a negative remanence in the low-field
part of the \AIAO ordered phase [Fig.~\ref{fig2}(a,b)]. Whereas several
pyrochlore compounds adopt the \AIAO structure \cite{Lhotel_2015,
  Xu_2015-1,Bertin_2015,Tian_2016,Ishikawa_2012,Yamaura_2012} and form domains
\cite{Ma_2015, Tardif_2015, Opherden_2017-1, Hirose_2017}, \NHO is the first
\AIAO antiferromagnet where such a negative remanent magnetisation has been
observed. The underlying all-in-all-out order in \NHO is established below
$T_{\mathrm{N}}$~=~0.48 K and is stable for external magnetic fields up to
$\mathrm{\mu}_0 H_{dc}$~=~0.27~T [Fig.~\ref{fig3} and Fig.~\ref{fig:supp_fig2}].

\begin{figure}[h!]
  \includegraphics[width=.99\columnwidth]{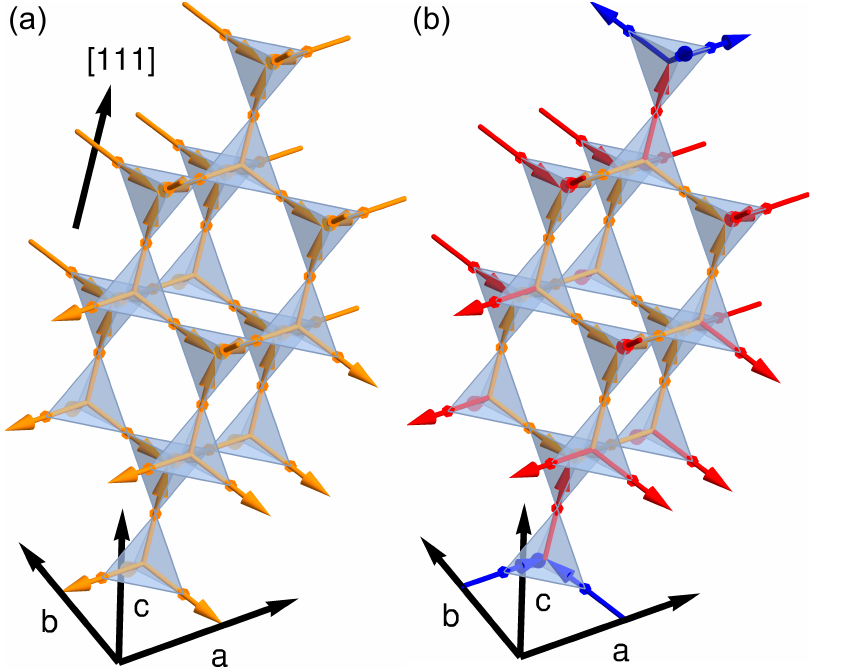}
	\caption{Pyrochlore structure formed by a lattice of
    corner-sharing tetrahedra and \AIAO magnetic order. (a) All-in--all-out ordered state
    in the pyrochlore structure with a,b,c crystal axes and [111] direction
    indicated. All spins are oriented along their local $\langle111\rangle$
		direction pointing either into or out from
		the centre of their tetrahedra. Shown is the AIAO state, the AOAI state (not
		shown) is obtained by flipping all spins. (b) Spherical AIAO domain in AOAI 
		background oriented along [111] resulting in a negative magnetisation. Blue (orange) spins belong to AOAI (AIAO) bulk tetrahedra, red spins
		are boundary spins of the AIAO domain.
		\label{fig:pyrochlore_structure}}
\end{figure}
\NHO belongs to the class of cubic pyrochlore oxides of the composition
$R_2T_2$O$_7$ where $R$ is typically a trivalent rare-earth ion and $T$ a
tetravalent transition-metal ion. In these structures the $R$ and the $T$ ions
both form a sublattice of corner-sharing tetrahedra
[Fig.~\ref{fig:pyrochlore_structure}]. An asymmetrical arrangement of eight
bivalent oxygen ions around each $R$ ion leads to a strong crystal electric
field (CEF) splitting of their $J$+1 multiplet by several hundred Kelvin
\cite{Gardner_2010}. The resulting CEF ground state is a pseudo spin-half state
with a strong local anisotropy and the spins are forced to point along the
corner-to-center direction of each tetrahedron corresponding to the four
equivalent local $\langle$111$\rangle$ directions of the cubic lattice. In the
\AIAO state all spins in a tetrahedron either point in or out resulting in two
distinct realisations: a fourth of the spins are oriented either parallel (AIAO)
or antiparallel (AOAI) to the $[111]$ axis. This allows for the formation of
domains which we argue to be central to the experimental observations detailed
next.

\sectionn{Results}
\begin{figure*}[t]
  \includegraphics[width=\linewidth]{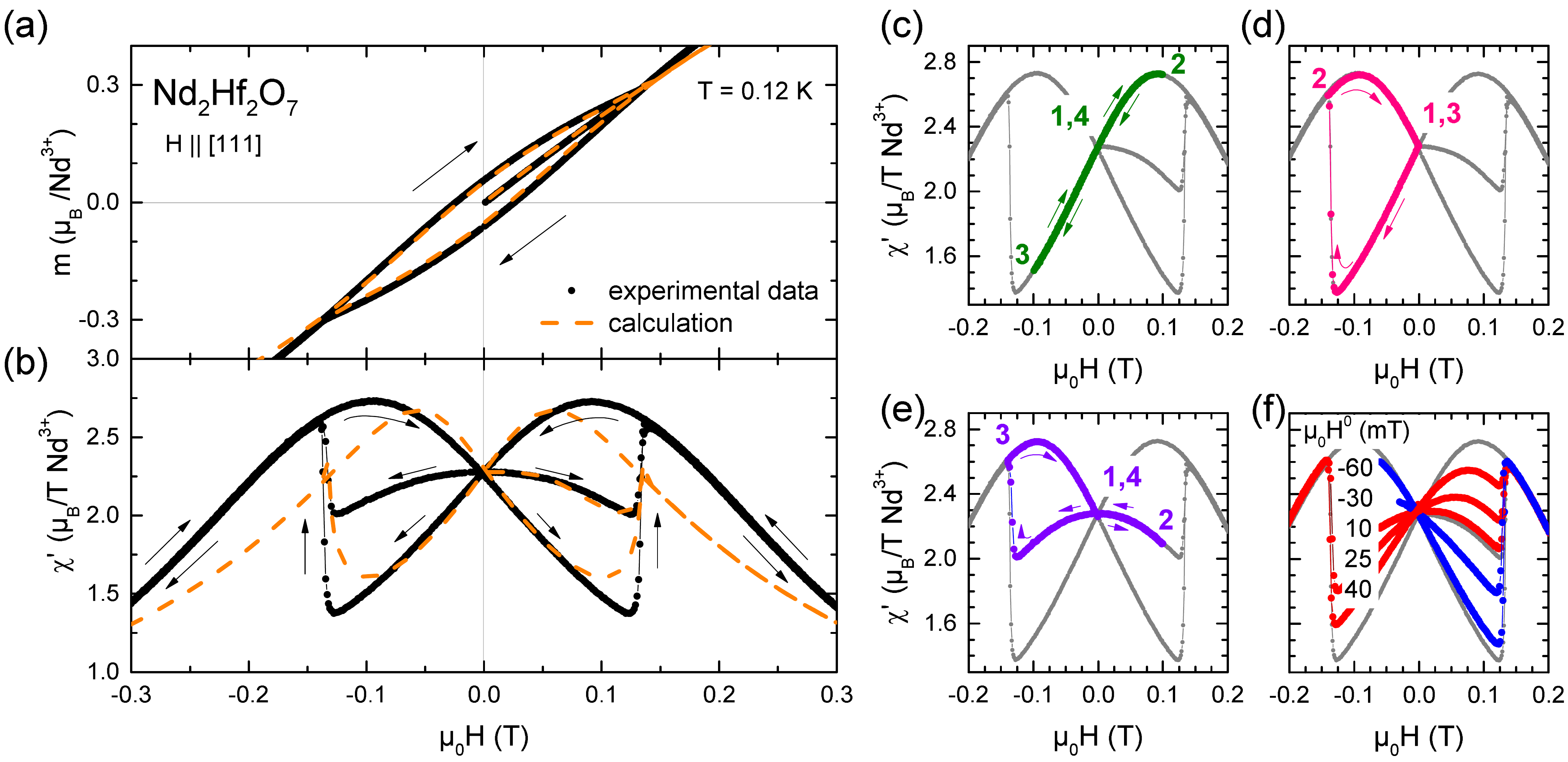}
  \caption{Magnetic-field dependence of the dynamic susceptibility. (a)
    Experimental results (black line) and results from calculation (orange
    dashed line) of the magnetization (a) obtained by integrating the dynamic
    susceptibility (b), both for $H\|$[111] at a temperature of 0.12 K . For
    fields smaller than $\pm125$~mT ($\pm H^{\ast}$) the susceptibility shows an
    inverted hysteresis. For higher fields a sharp transition to a polarised
    domain state is visible. (c) Staying below the critical field $H^{\ast}$ the
    state is stable (d,e) The state can be changed by applying a high positive
    or negative field or by applying a heat pulse. (f) AC susceptibility for
    different applied fields $H^0$ while cooling the sample below
    $T_{\mathrm{N}}$.
    \label{fig2}}
\end{figure*}
The hysteresis behavior shows the following salient properties: There is (a) an
inverted hysteresis loop with a large negative remanent magnetisation, (b) which
is stable, reproducible and fully tunable by cooling the sample in a finite
field, and (c) can be reset by applying a heat pulse above $T_{\mathrm{N}}$;
finally, (d) it is highly anisotropic, only occurring for specific field
directions.

The fully inverted hysteresis loop of \NHO observed for $H\|$[111]
[Fig.~\ref{fig2}] covers a fixed area and appears only if the external magnetic
field exceeds a certain temperature dependent value $H^{\star}(T)$ (up to 0.14 T
at 0.12~K for $H\|$[111]) where the susceptibility shows a sharp and pronounced
jump [Fig.~\ref{fig2}(b)].

In contrast, for fields oriented along the [001] direction we find no hysteresis
[Fig.~\ref{fig:supp_fig2}(b)]. At the same temperature $H^{\star}$ is
significantly smaller than the critical field $H_c$ at which the transition
occurs for $H\|$[001] [Fig.~\ref{fig:supp_fig2}(c)]. Further, it strongly
depends on temperature for $T \leq T_{\mathrm{N}}$ [Fig.
\ref{fig:supp_ACS_Hpar111_diffTemp}]. For temperatures higher than
$T_{\mathrm{N}}$, no hysteresis is observed.

We find a large coercivity, about 24~mT at a temperature of 0.12~K, to be
compared to recent findings in ferromagnetic thin films which show a reversed
hysteresis with a coercive field of only about 2~mT \cite{Maity_2017}. Also the
remanence is unusually large at 0.06~$\mu_\mathrm{B}$/Nd$^{3+}$, 100 times
larger than (the conventional 'non-inverted' hysteresis) in Cd$_2$Os$_2$O$_7$
(up to $6\cdot 10^{-5} \mu_\mathrm{B}$/Os) \cite{Hirose_2017}.

Next, we demonstrate that the sample adopts distinct stable states depending on
the field/temperature history. We start by considering the stability of the
observed hysteresis curves to magnetic field changes. In Fig.~\ref{fig2}(c,e) we
show that when staying below $H^\star$ the magnetization is free of hysteresis
and the state is perfectly stable. The susceptibility follows a unique perfectly
reproducible trajectory when sweeping the magnetic field back and forth staying
below $H^{\star}$, both when in the polarised state [Fig.~\ref{fig2}(c), green
data points] as well as in the unpolarised state [Fig.~\ref{fig2}(e), violet
data].

Moreover, one can reliably switch between the different states of the system. In
Fig.~\ref{fig2}(d,e) we show such a protocol. Starting in the positive polarised
or unpolarised state sweeping the field below $-H^{\star}$ switches the system
to the negative polarised state in which the susceptibility follows a distinct
stable trajectory. Then applying a thermal cycle above the critical temperature
after switching off the field allows to return the sample to the unpolarised
state.

Additionally, the observed hysteresis in the susceptibility may be continuously
and fully tuned. By cooling the sample below $T_{\mathrm{N}}$ in a finite
magnetic field $|H^\mathrm{0}| < |H^{\star}|$, we prepare a family of distinct
stable states with associated susceptibility curves [Fig.~\ref{fig2}(f)]. The
observed trajectories continuously interpolate between the maximally polarised
conditions. These states are again perfectly stable to changes in the magnetic
field below the critical field with a unique trajectory defined by the initially
prepared state. This is in stark contrast to the known behavior of ferromagnetic
materials where a field change or temperature variation within the ordered state
leads to domain wall orientation, hence, a reduced hysteresis loop and different
trajectories of $m(H,T)$.
\begin{figure*}[t]
  \includegraphics[width=\linewidth]{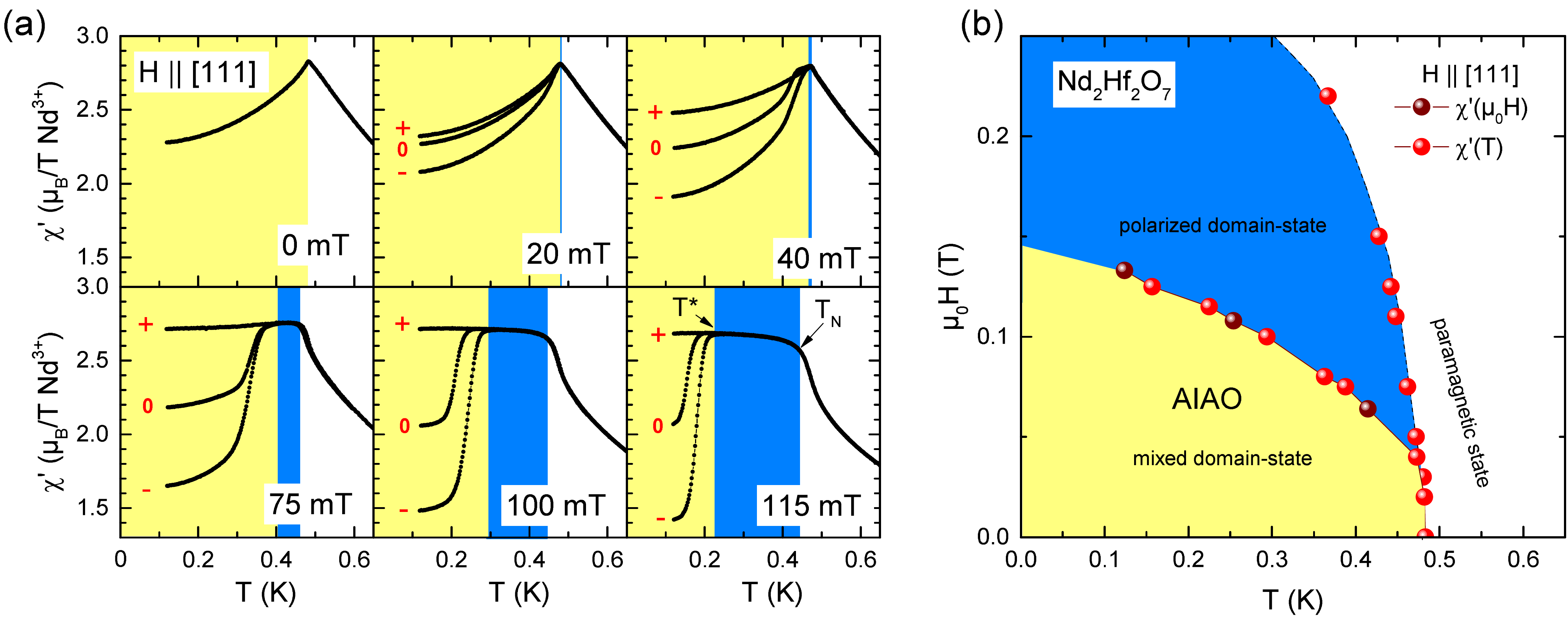}
  \caption{(a) Temperature dependence of the dynamic susceptibility of \NHO for
    magnetic DC fields applied along $H\|$[111]. The sample was prepared to be
    either in the positive polarised state (+), the negative polarised state (-)
    or the unpolarised state (0). The blue shaded area denotes the respective
    temperature range of the single domain state. (b) Phase diagram for \NHO for
    $H\|$[111]. The phase boundary of the \AIAO order was determined by the
    kinks/peaks of $\chi^\prime (T)$ (red spheres). The border between the mixed
    domain phase and the polarised domain phase was constructed by taking the
    temperature at each field (red spheres) below which the AC
    susceptibility starts to differ when polarizing in negative compared to
    positive external fields $|H| > H_c$ and the field strength of the end points
    of the hysteresis in the dynamic susceptibility (brown spheres). The thin black dashed line is a guide
    to the eye.
    \label{fig3}}
\end{figure*}

Finally, we demonstrate that the AIAO ordered state of \NHO splits into two
distinct phases as seen in the AC susceptibility for external magnetic fields
applied along [111] [Fig.~\ref{fig3}] as well as [110]
[Fig.~\ref{fig:supp_ACS_Hpar110_diffFields}]. At low fields and temperatures the
system can adopt different states whereas at high fields a unique state appears.

For each field, the sample was cooled under an overcritical positive field (+),
an overcritical negative field ($-$) and in zero field (0). At low temperatures
the susceptibility of these states differs: the negatively polarised state
showing a lower susceptibility than the positively polarised state and the
unpolarised (zero field cooled) state being intermediate the polarised states.

For a sufficiently strong field we find a new phase setting in at larger
temperatures where the susceptibility for the differently prepared states
coincides before diverging at lower temperatures. The entrance into this unique
state is accompanied by a sharp change of $\chi^\prime$ for the negatively and
non-polarised states and the susceptibility only depends weakly on temperature
within this phase. The temperature range where this state exists increases at
increasing external field (blue shaded areas in Figs.~\ref{fig3} and
\ref{fig:supp_ACS_Hpar110_diffFields}). At an even higher temperature,
$\chi^\prime (T)$ exhibits a peak (kink for higher fields) and the
antiferromagnetic order is broken.

We construct the field-temperature phase diagram for $H\|$[111] [Fig.
\ref{fig3}(b)] by the temperature of the kink as well as by the endpoints of the
hysteresis in the temperature and field dependence of the AC susceptibility
[Fig. \ref{fig3} and Fig. \ref{fig:supp_ACS_Hpar111_diffTemp}]. Whereas the kink
represents the phase boundary to the all-in-all-out ordered state, the latter
criteria separates the mixed domain state from the polarised domain state. In
the Supplementary the corresponding phase diagrams for different field
directions are provided.

\sectionn{Discussion} We interpret these results as originating from domains of
the two possible realisations of the \AIAO order. At strong fields only the
fully polarised domain state exists, whereas within the hysteretic part of the
\AIAO state, the sample can contain any mixture of AIAO and AOAI domains.
Preparing the sample in a finite field allows for tuning the ratio of AIAO to
AOAI domains up to the fully polarised single domain state. The results then
suggest that within each phase these domain configurations are stable to changes
in temperature and magnetic field.

We propose an explanation of the anisotropic and inverted hysteresis and the
negative remanent magnetization via the preferential formation of oppositely (to
the magnetic field) polarised domain walls present within a non-magnetic bulk
background phase. This explains (a) the negative remanence, (b) the tunability
of the hysteresis via cooling in a finite field, (c) the resetting of the sample
state via a heat pulse and (d) the anisotropy.

We model the system via a free energy for a mixture of two bulk phases with
different energies given by
\begin{align}
  f &=  x \left[ E_{\mathrm{AIAO}}(H)- E_{\mathrm{AOAI}}(H)\right]+x(1-x) \left[E_{\mathrm{DW}}- m_{\mathrm{DW}} H \right] \notag\\
    &+T \left[ x \ln(x) +(1-x) \ln(1-x)\right] \, ,
\end{align}
with the volume fraction $x$ of the AIAO bulk phase, the corresponding bulk
energies $E_{\mathrm{AIAO/AOAI}}$, a contribution of the domain walls
proportional to the surface between phases $x(1-x)$ with associated cost
$E_{\mathrm{DW}}$ due to broken bonds and magnetisation $m_{\mathrm{DW}}$, and
the mixing entropy between phases proportional to the temperature $T$.

For the comparison with the experimental hysteresis curves in
Fig.~\ref{fig2}(a,b) we use the energy and magnetisation of the bulk phases
obtained from a classical treatment of the microscopic hamiltonian which
reproduces the magnetisation behavior observed for magnetic fields oriented
along [100] (see Supplementary). We emphasise that we do not fit parameters to
the hysteric data, but rather use the parameters obtained from the non-hysteric
magnetisation curve observed for a different field direction and obtain good
agreement with the hysteric experimental data.

This model is based on the following key observations: The \AIAO state can be
realized by two configurations, related to each other by flipping all spins on
their Ising axis [Fig.~\ref{fig:pyrochlore_structure}]. Within the \AIAO order
one fourth of the spins is parallel (AIAO configuration) or antiparallel (AOAI
configuration) to the [111] direction. An effective canting of the spins out of
their local Ising-anisotropy axis, as a result of the dipolar-octupolar nature
of the Nd$^{3+}$ ions, can lead to a gain of Zeeman energy which differs between
these two configurations depending on the field direction
\cite{Opherden_2017-1}. In particular, for $H\|$[111] the AIAO and AOAI states
are split in energy. Thus, we model the system as a mixture of two bulk phases
with an energy difference dependent on the applied magnetic field. We note that
at zero field these bulk phases carry no magnetisation and are degenerate.

In contrast, for a field applied parallel to the [001] direction, the projection
of the field onto the four $\langle$111$\rangle$ directions is equal. Therefore,
the field does not distinguish between the AIAO/AOAI states, the distribution of
domains does not change when applying a field and, thus, the susceptibility is
free of hysteresis.

At finite temperature both phases will be present to maximise the entropy which
we model by including a mixing entropy in the free energy. Domain walls between
the phases cost an energy $E_{\textrm{DW}}$ due to the broken bonds at the
boundary [Fig.~\ref{fig:pyrochlore_structure}(b)].

To explain the negative remanence we note that domain walls carry a
magnetisation $m_{\textrm{DW}}$; and that this magnetisation preferentially
opposes the applied field. This happens because walls between AIAO and AOAI
domains with a negative magnetisation are kinematically favoured as they have a
lower energy barrier than walls with a positive magnetisation (see
Supplementary). In consequence, once domains have formed at low temperature and
finite field, the sample retains a negative remanent magnetization after
removing the field: in zero field, both AIAO and AOAI domains have zero
magnetisation, so that only that of the domain walls remains. We find that
spherical domains carry a magnetization along one of the four equivalent
$\langle$111$\rangle$ directions [Fig.~\ref{fig:pyrochlore_structure}(b) and
Supplementary].

Naturally, the magnetization of a random collection of domain walls vanishes.
Thus, applying a heat pulse $T > T_{\mathrm{N}}$ to the sample allows a
relaxation of the domains in which all symmetry equivalent domains are created
equally, and the magnetization vanishes again.

By contrast, when preparing the sample in a partially polarised state by cooling
down in a finite field, both AIAO and AOAI domains are present with a ratio
determined by the field strength and tunable from fully positive to fully
negatively polarised. The hysteresis thus follows the weighted average of the
fully positively and fully negatively polarised sample states.

\sectionn{Conclusions} Our findings constitute, to the best of our knowledge,
the first observation of an inverted hysteresis/negative remanence in a bulk
antiferromagnet. The mechanism -- a nonequilibrium popoulation of negatively
polarised domain walls -- appears entirely novel. It invites more detailed
investigations, in particular with regard to the nucleation process leading to
the disappearance of the phenomenon, and the spatial distribution of the domain
walls themselves.

By contrast, previous instances of negative remanence were essentially
ferri-magnetic in nature, in that they relied on the non-cancellation of moments
of inequivalent ferromagnetic subsystems, either in form of thin films \cite{Maity_2017} or
different ionic species \cite{Ohkoshi_2001}.

Not only is our hysteresis loop highly reproducible and stable to magnetic field
and temperature changes, but also exquisitely tunable. We demonstrate precise
control of the system's response via the preparation of a continuous family of
distinct non-equilibrium states.

Thus, \NHO provides a non-equilibrium landscape controllable via small magnetic
fields or temperature pulses. Controlled tunability of magnetic structures
underpins the field of spintronics, and currently the search for novel types of
magnetic structures and their manipulation is a central theme there. The
preferential coupling in \NHO of a weak field to domain walls provides an unsual
handle on this unconventional magnet, while the antiferromagnetic domains
themselves do not interact with each other via a bulk magnetisation. It is
further promising that \NHO may be prepared as epitaxially grown thin films of a
few nanometer thickness \cite{Wei_2009}. Each of these items is desirable for
technological applications underlining the potential for inclusion in submicron
devices.


\bibliography{References}

\subsection*{Acknowledgement}
We acknowledge the Helmholtz Gemeinschaft for funding via the Helmholtz Virtual
Institute (Project No. VH-VI-521) and DFG through SFB 1143. We also acknowledge
support by HLD at HZDR, member of the European Magnetic Field Laboratory (EMFL).

%
%

\clearpage
\section{Supplementary Material}
\setcounter{equation}{0} \setcounter{figure}{0} \setcounter{table}{0}
\setcounter{page}{1} \makeatletter
\renewcommand{\theequation}{S\arabic{equation}}
\renewcommand{\thefigure}{S\arabic{figure}} \renewcommand{\bibnumfmt}[1]{[S#1]}

\subsection{Experimental protocol}
\NHO single crystalline samples were investigated by means of dynamic (AC)
susceptibility using a pair of compensated coils with frequencies between 11~Hz
and 11~kHz and field amplitudes, $H_{\mathrm{ac}}$, between 1.8 and
9.6~$\mathrm{\mu}$T, and static (DC) magnetization at a field amplitude
$H_{\mathrm{dc}}$ of 0.1~T using a commercial vibrating sample magnetometer. The
AC-field direction was aligned parallel to the DC field. If not stated
differently, the AC susceptibility was measured at $f$ = 1111~Hz. The single
crystals were grown by the floating-zone technique using a high-temperature
optical furnace.
\subsection{Influence of the AC frequency}
\begin{figure}[h!]
	\centering \includegraphics[width=.99\linewidth]{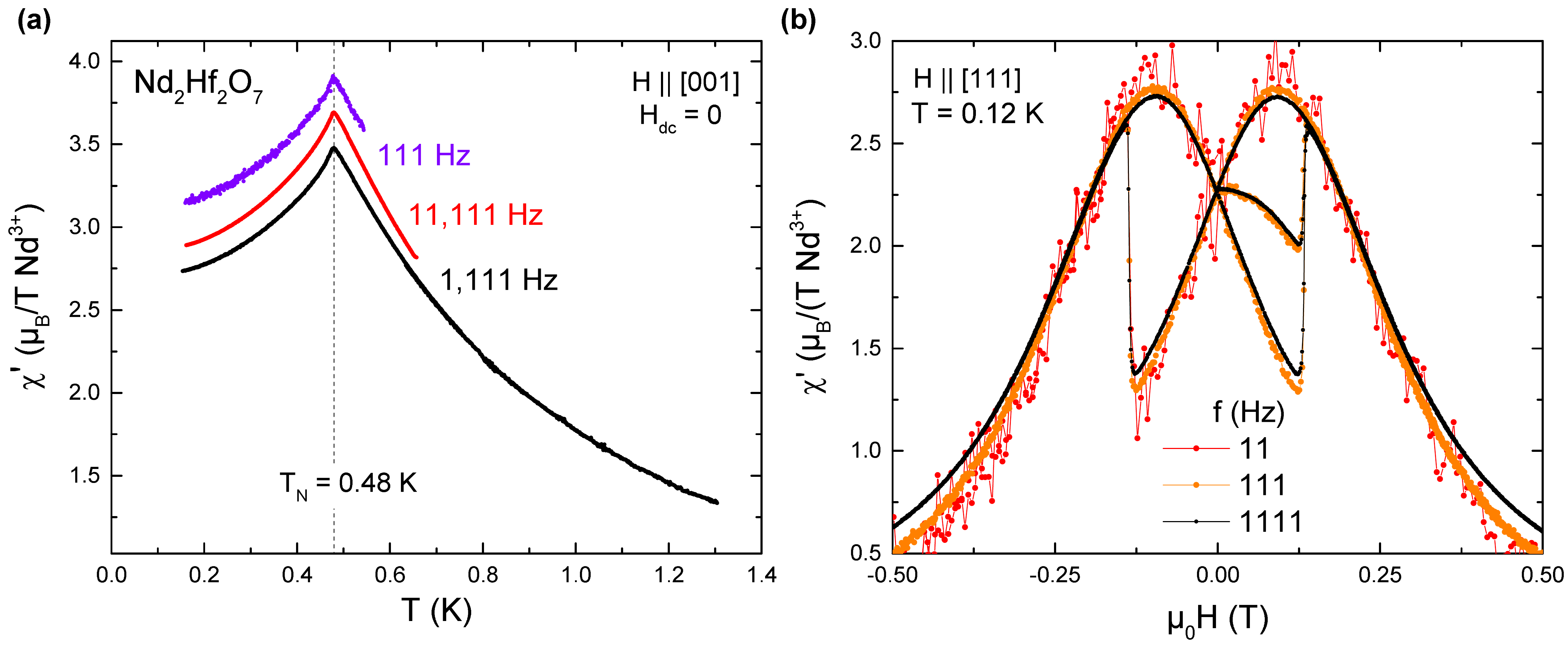}
	\caption{Temperature (a) and magnetic field (b) dependence of the dynamic
    susceptibility without an external DC magnetic field for different
    frequencies of the AC magnetic field. \label{fig:supp_ACS_diffFreq}}
\end{figure}
The influence of the frequency on the dynamic susceptibility for \NHO was
investigated by measuring the temperature dependence of $\chi_\mathrm{ac}$ for
zero DC magnetic field and $H_\mathrm{ac}\|$[001][Fig.
\ref{fig:supp_ACS_diffFreq} (a)] as well as the field dependence for an external
field along [111] at 0.12 K [Fig. \ref{fig:supp_ACS_diffFreq} (b)].

In both cases the measured dynamic susceptibility does not depend on the used AC
frequency over four orders of magnitude. Therefore we conclude that even at the
highest measured frequency the spin system can still respond fully to the
applied AC field and, thus, the dynamic susceptibility behaves in this frequency
range as the static susceptibility $\chi_\mathrm{dc}$. In consequence, the
magnetization of \NHO can be obtained by integrating the dynamic susceptibility
\begin{equation}
  M \sim \int \chi_\mathrm{ac} \, \mathrm{d}H.
\end{equation}

\subsection{Results for H $\|$ [001]}
\begin{figure}[t]
  \centering \includegraphics[width=\linewidth]{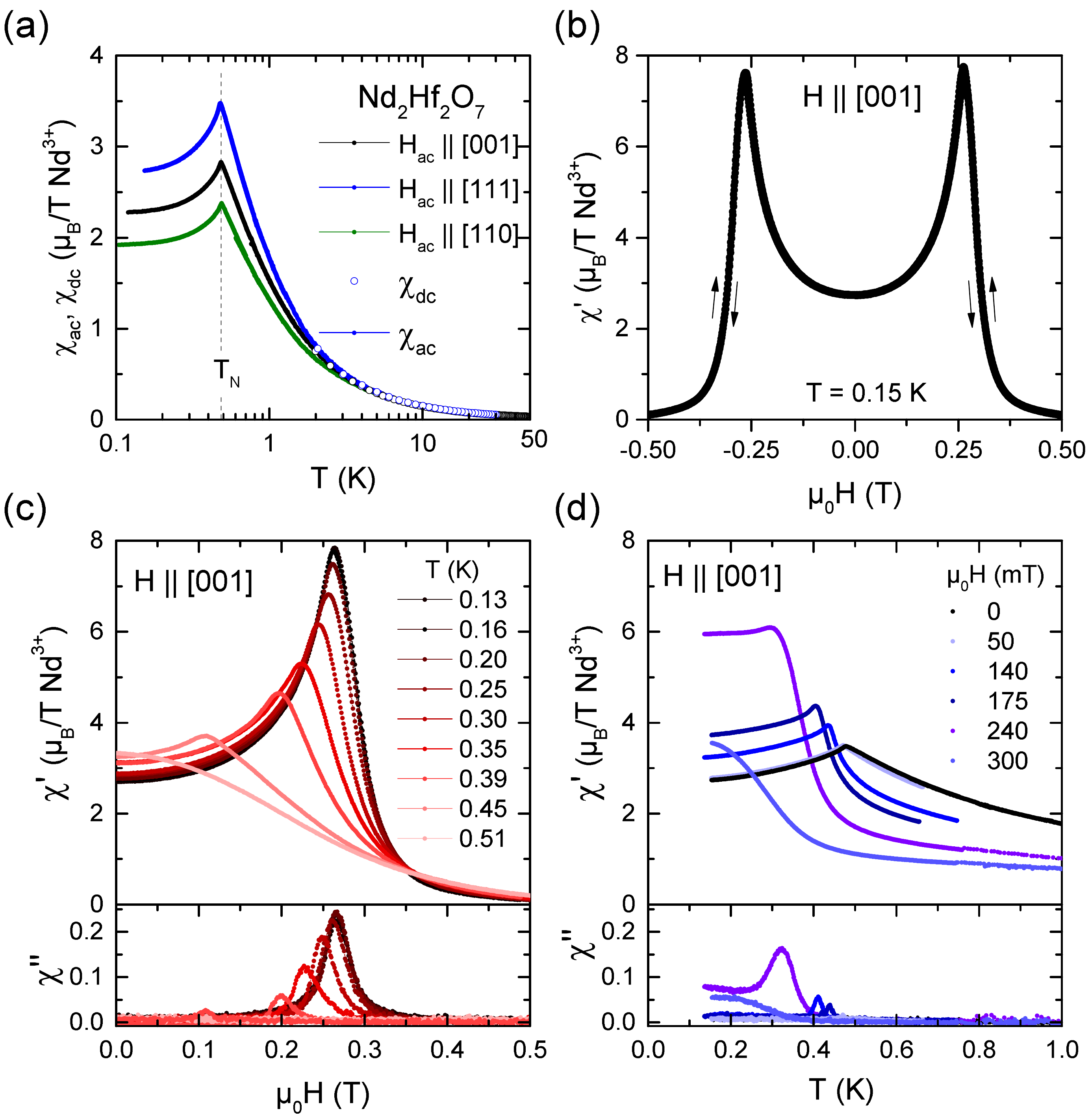}
  \caption{Dynamic susceptibility of \NHO. (a) $\chi_\mathrm{ac}^\prime$ without
    an external magnetic field and $\chi_\mathrm{dc} = M/H$. (b) Field
    dependence of $\chi^\prime$ for $H\|$[001] at a temperature of 0.15 K for a
    full hysteresis cycle. (c) Field dependence of $\chi^\prime$ and
    $\chi^{\prime\prime}$ for $H\|$[001] for various temperature. (d)
    Temperature dependence of $\chi^\prime$ and $\chi^{\prime\prime}$ for
    $H\|$[001] for various applied magnetic fields.}
  \label{fig:supp_fig2}
\end{figure}
The dynamic susceptibility of \NHO in zero DC field follows a Curie-Weiss law
above $T_{\mathrm{N}}$ and coincides with the static susceptibility
($\chi_\mathrm{dc} = M/H$).

At a temperature of $T_{\mathrm{N}}$ = 0.48 K a cusp is visible [Fig.
\ref{fig:supp_fig2}(a)]. This is a characteristic feature of the phase
transition to the antiferromagnetically ordered all-in-all-out state
\cite{Opherden_2017-1, Lhotel_2015} as recently confirmed for \NHO by neutron
scattering and $\mu$SR experiments \cite{Anand_2017, Anand_2015}. (The ordering
temperature was determined to be slightly higher in polycrystalline samples.)

The phase transition can also be seen in the field dependence of the AC
susceptibility for an external magnetic field along [001] where $\chi^\prime$
shows a pronounced maximum and no sign of hysteresis
[Fig.~\ref{fig:supp_fig2}(b)]. The maximum in $\chi^\prime$ is accompanied by a
peak in $\chi^{\prime\prime}$, centred at the same field $H_c$.

The intensities of both peaks are reduced if the temperature is increased
towards $T_{\mathrm{N}}$ and $H_c$ is lowered [Fig.~\ref{fig:supp_fig2}(c)].
Above $T_{\mathrm{N}}$, the susceptibility shows no maximum.

From this data the phase boundary of the all-in-all-out phase for $H\|$[001] was
determined [black dots in Fig.~\ref{fig:supp_phase_diagrams}(a)]. Further points
of the phase boundary were obtained by measuring the temperature dependence in
presence of an external DC field [Fig.~\ref{fig:supp_fig2}(d)] and analysing the
peak temperature of $\chi^{\prime\prime}$ [grey dots in
Fig.~\ref{fig:supp_phase_diagrams}(a)].

\subsection{Behavior of the [111]-hysteresis for different temperatures}
\begin{figure}[h!]
	\centering
  \includegraphics[width=0.9\linewidth]{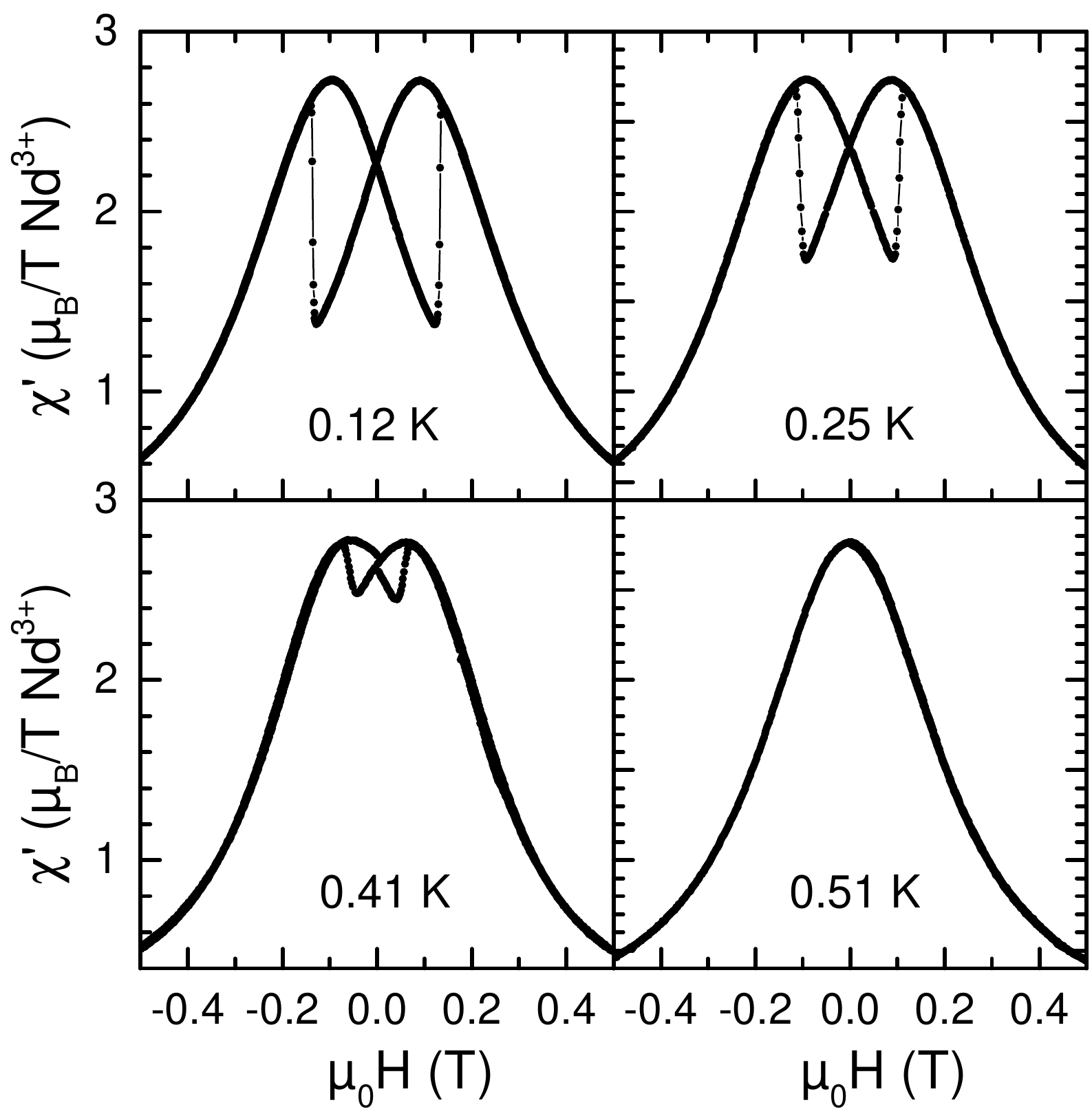}
	\caption{Magnetic field dependence of the dynamic susceptibility for different
    temperatures and $H\|$[111].}
	\label{fig:supp_ACS_Hpar111_diffTemp}
\end{figure}
The critical field below which a hysteresis is observed in the dynamic
susceptibility of \NHO for $H\|$[111] is reduced at larger temperatures
(Fig.~\ref{fig:supp_ACS_Hpar111_diffTemp}). Whereas the difference between the
positively and negatively polarised state within the multiple domain phase at a
constant temperature remains nearly unchanged, the field width of the hysteresis
is decreasing.

Furthermore, the width of the transition also increases with increasing
temperature. If the temperature exceeds the N\'{e}el temperature
$T_{\mathrm{N}}$ of 0.48 K the hysteresis vanishes completely.
\subsection{Hysteresis for H $\|$ [110]}
\begin{figure}[h!]
	\centering
  \includegraphics[width=.99\linewidth]{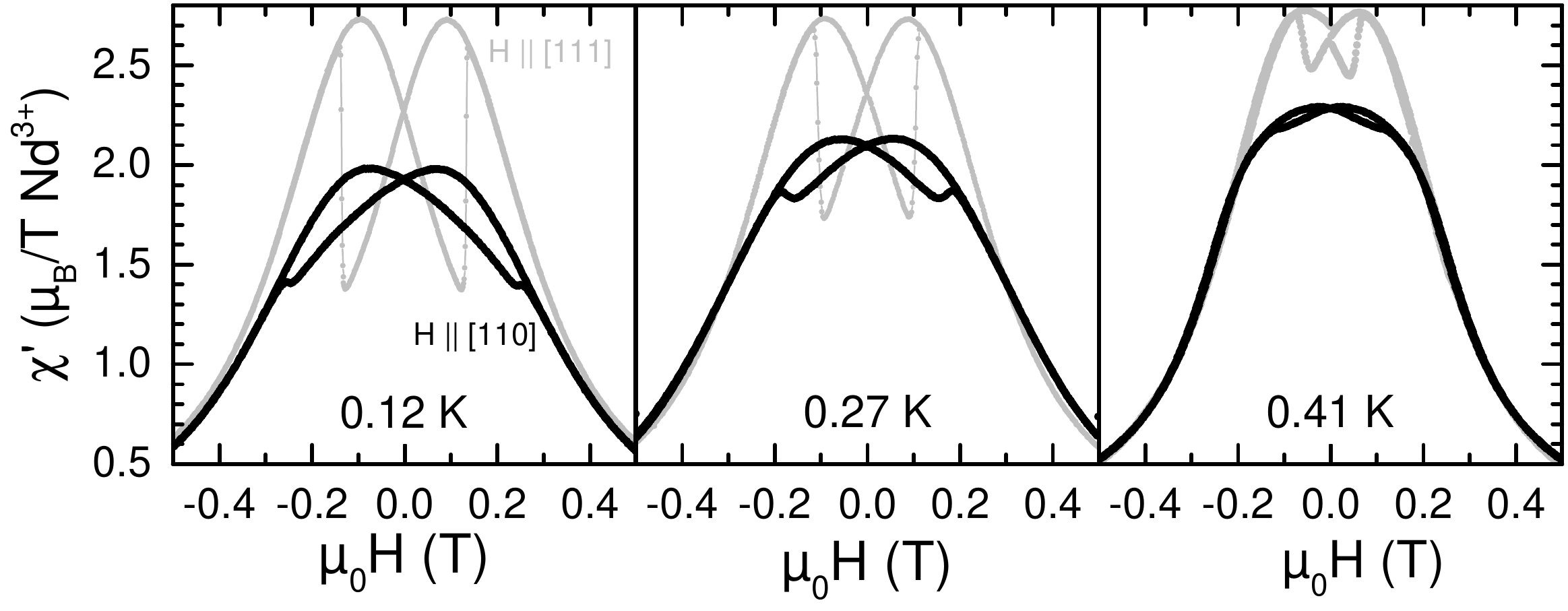}
	\caption{Comparison of the AC susceptibility hysteresis for an applied field
    parallel to [110] (black) and [111] (grey) for different temperatures.}
	\label{fig:supp_ACS_Hpar110_diffTemp}
\end{figure}
For an applied field along [110] an inverted hysteresis loop is observed as well
[Fig.~\ref{fig:supp_ACS_Hpar110_diffTemp}].

\begin{figure}[h!]
	\centering
  \includegraphics[width=.99\linewidth]{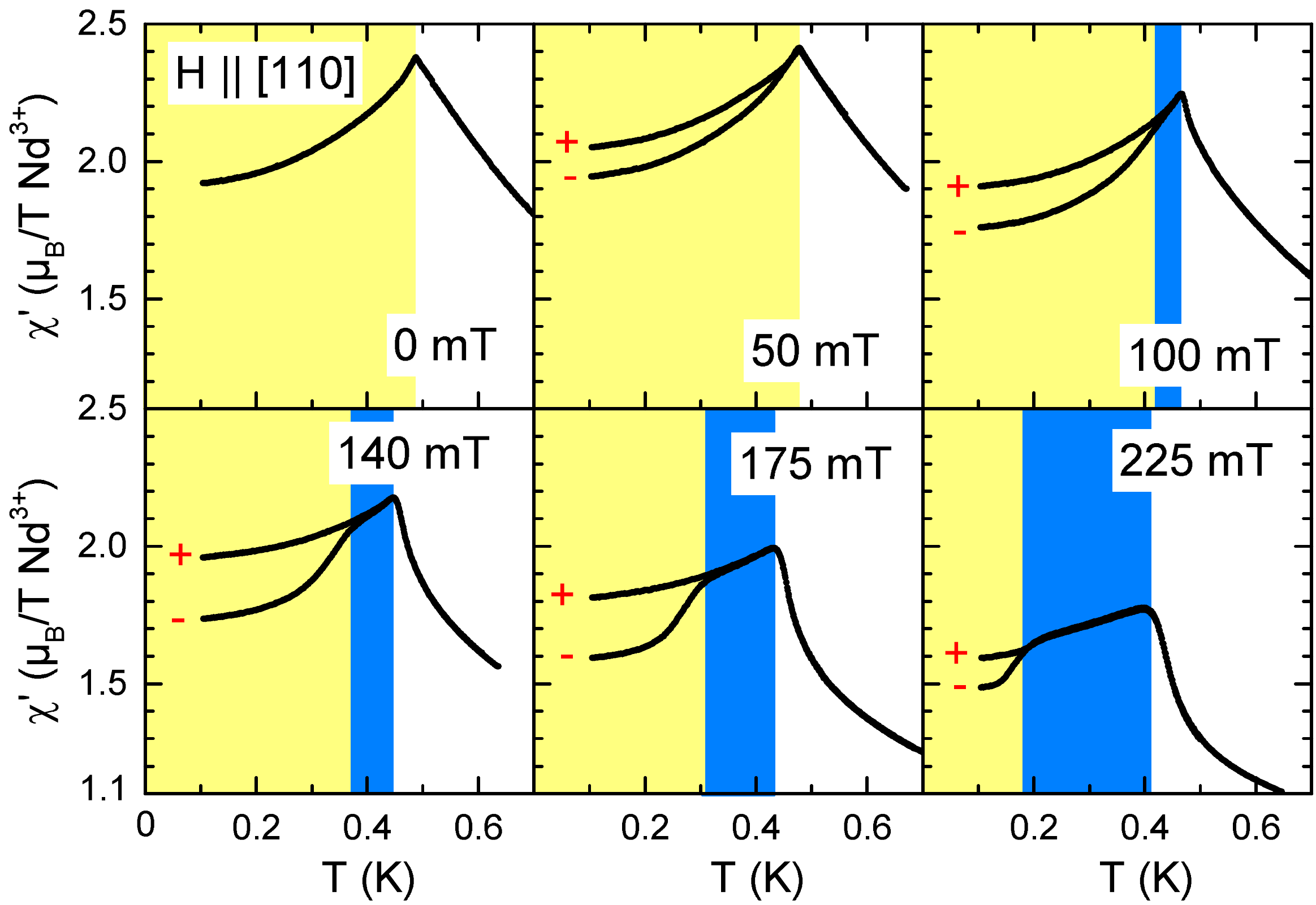}
	\caption{Temperature dependence of the dynamic susceptibility for different
    preparation fields in the orientation $H\|$[110]. The sample was prepared to
    be either in the positively polarised state (+) or the negatively polarised
    state (-). The blue shaded area denotes the respective temperature range of
    the single domain state.}
	\label{fig:supp_ACS_Hpar110_diffFields}
\end{figure}

Whereas the hysteresis extends over a larger field range
(Fig.~\ref{fig:supp_ACS_Hpar110_diffTemp}) the difference between the positive
and negative polarised states is smaller than for a field along [111] (see
Fig.~\ref{fig3}(a) and Fig.~\ref{fig:supp_ACS_Hpar110_diffFields}).
\subsection{Comparison of all main field directions}
\begin{figure}[h!]
	\centering \includegraphics[width=0.9\linewidth]{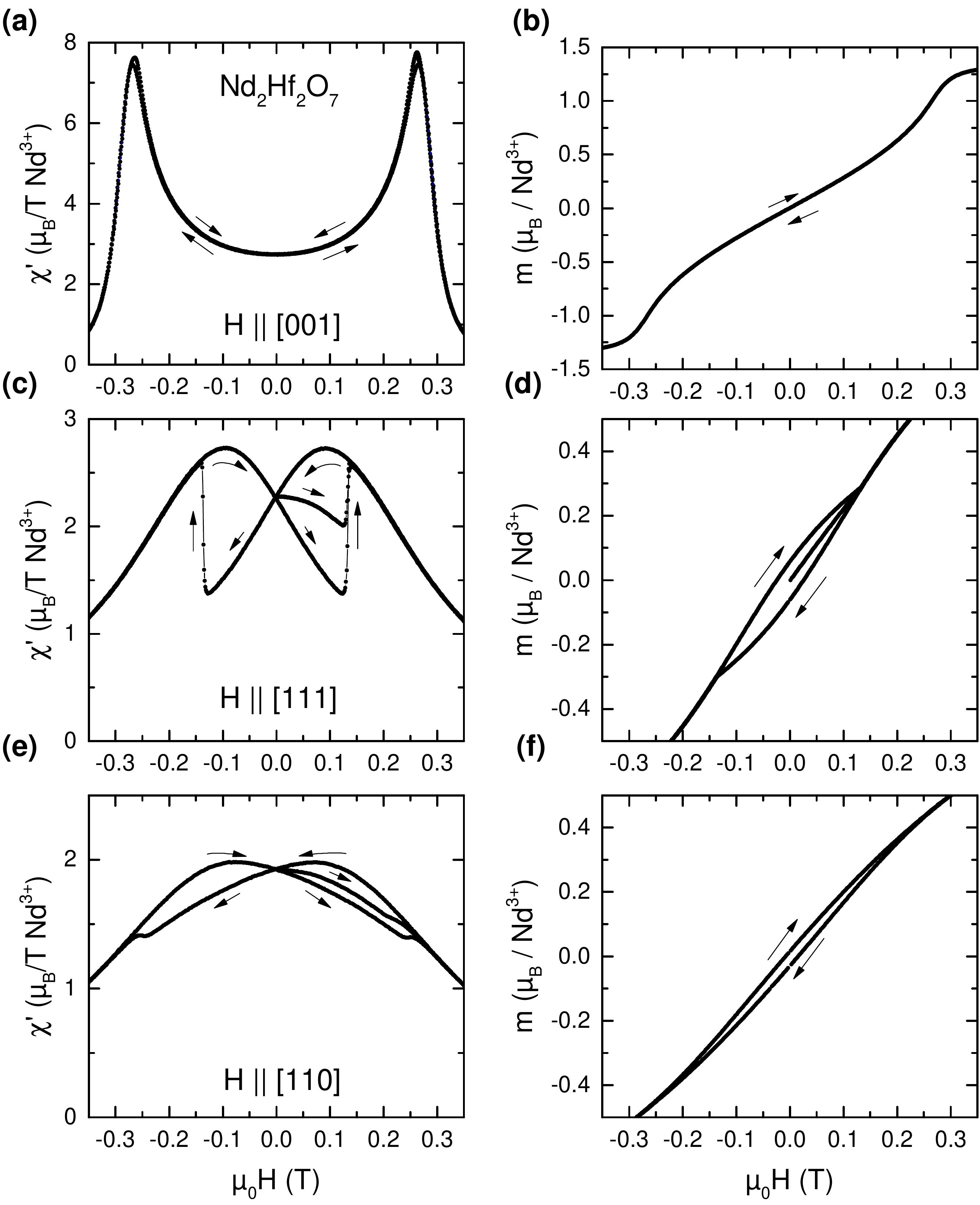}
  \caption{Magnetic field dependency of the dynamic susceptibility (left panels)
    and magnetization (right panels) for $H\|$[001] (a,b), $H\|$[111] (c,d) and
    $H\|$[110] (e,f), obtained by integration, at a temperature of 0.12 K, 0.15
    K and 0.11 K.}
	\label{fig:supp_ACS_Hpar111_120mK_integratedMagComparison}
\end{figure}

The strong anisotropic behavior of \NHO can be most clearly seen in the direct
comparison of all three main directions (Fig
\ref{fig:supp_ACS_Hpar111_120mK_integratedMagComparison}).

While for $H\|$[001] the phase transition has the most drastic effect on the
dynamic susceptibility, no hysteresis can be observed [panels (a) and (b)].

In contrast both for $H\|$[111] [panels (c) and (d)] and for $H\|$[110] [panels
(e) and (f)] an inverted hysteresis loop is found. In comparison the hysteresis
loop for $H\|$[111] extends over a smaller field range, but shows a larger
response than for $H\|$[110].

\begin{figure}[h!]
	\centering
  \includegraphics[width=.99\linewidth]{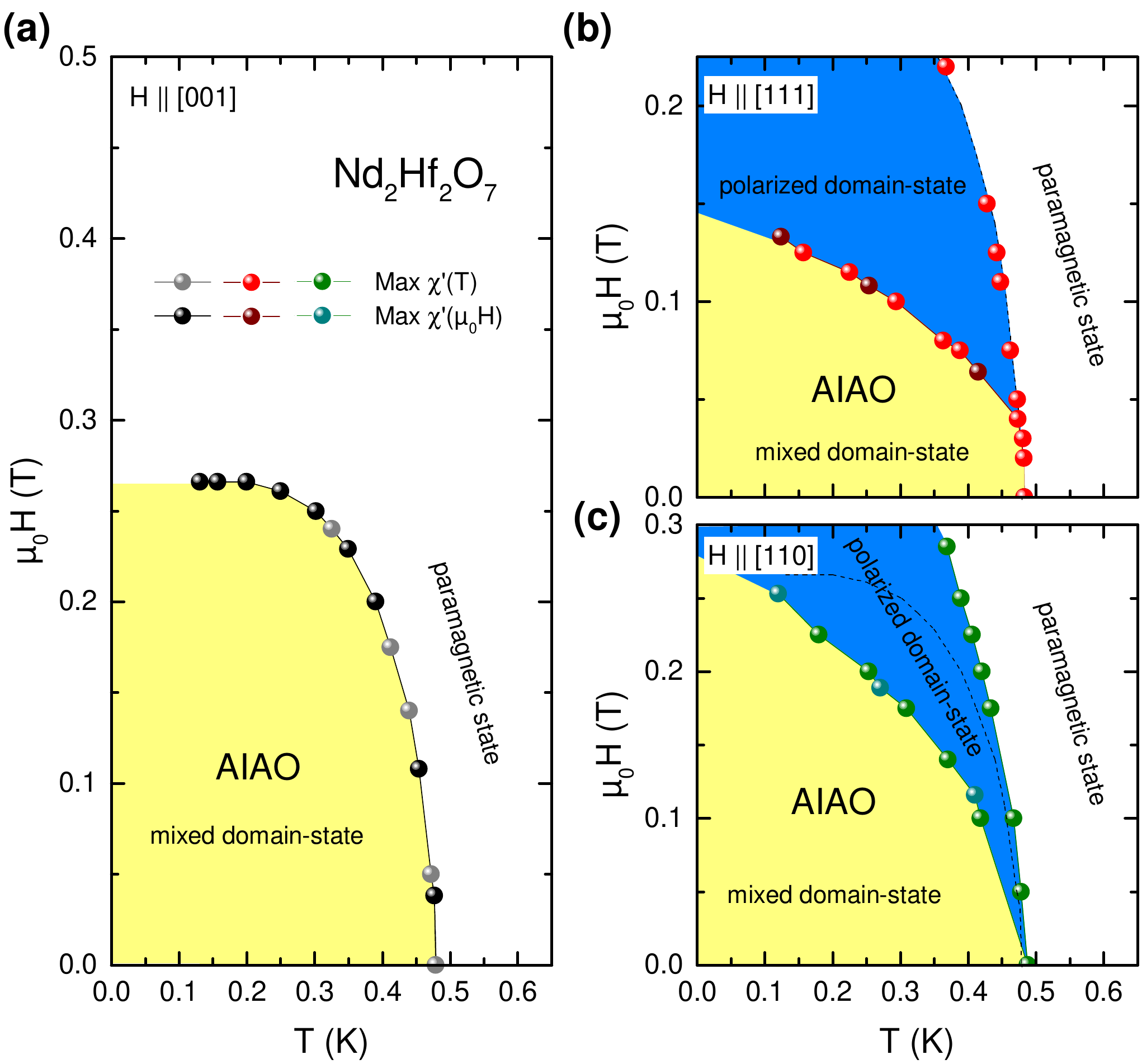}
	\caption{Comparison of the phase diagrams of \NHO for different field
    directions, $H\|$[001] (a), $H\|$[111] (b) [same as main text Fig.~\ref{fig3}(b)], and
    $H\|$[110] (c).
    (a)  The phase boundary of the \AIAO order was determined by the
    maximum in the temperature (grey) and field (black) dependence of the
    dynamic susceptibility.
    (b,c) The phase boundary of the \AIAO order was determined by the
    kinks/peaks of $\chi^\prime (T)$ (red/green spheres). The border between the mixed
    domain phase and the polarised domain phase was constructed by taking the
    temperature at each field (red/green spheres) below which the AC
    susceptibility starts to differ when polarising in negative compared to
    positive external fields $|H| > H_c$ and the field strength of the end points
    of the hysteresis in the dynamic susceptibility (brown/grey spheres).
    The dashed black line in (b,c) is the phase boundary of the \AIAO order determined
    for $H\|$[001] in (a).\label{fig:supp_phase_diagrams}}
\end{figure}

In Fig~\ref{fig:supp_phase_diagrams} we compare the resulting phase diagrams for
the three main field directions.

As described above we obtain the phase-boundary of the hysteresis-free \AIAO
order for $H\|$[001] [panel (a)] from the temperature and field dependence of
the dynamic susceptibility shown in Fig.~\ref{fig:supp_fig2}.

Since for the other two field directions [panels (b) and (c)] we find hysteric
behavior we consider
the kink/peak in $\chi^\prime (T)$ [Fig.~\ref{fig3}(a) and
Fig.~\ref{fig:supp_ACS_Hpar110_diffFields}] to signal the transition into the
paramagnet. The transition line between the mixed and polarised domain phase was
constructed by determining the temperature at each field below which the AC
susceptibility shows a difference for different preparation fields
[Fig.~\ref{fig3}(a) and Fig.~\ref{fig:supp_ACS_Hpar110_diffFields}] and the end
points of the hysteresis in the dynamic susceptibility.

We note that the phase boundary of the \AIAO order agrees quite well between
$H\|$[001] and $H\|$[111], but differs strongly in the case of $H\|$[110] in
stark contrast to the results found in $\mathrm{Nd}_2\mathrm{Zr}_2\mathrm{O}_7$ \cite{Lhotel_2015}.

\subsection{Domain walls in the \AIAO phase}
In the main text we explain the observed anisotropic and inverted hysteresis and
the negative remanent magnetization via the preferential formation of domains
with boundaries which are polarised opposite to the applied magnetic field. This
model explains (a) the negative remanence, (b) the heat pulse ``effect'' , (c)
the mixing of states/partial preparation and (d) the anisotropy.

Here we provide additional details on the model underlying the \AIAO phase, the
different realizations of the \AIAO state, the resulting domains, specifically
the energetics/energy barriers favoring negatively polarised domain walls, and
propose a simple free energy capturing the magnetization hysteresis.

\subsection{All-in--all-out order}
Within the \AIAO order the spins on each sublattice point along their
corresponding Ising anisotropy axis, i.e. one of the symmetry-equivalent
$\langle 111 \rangle$ directions, with all spins belonging to a given
tetrahedron either pointing in towards or out from the center of the
tetrahedron. Thus, the \AIAO order can be realized by two configurations,
obtained from each other by flipping all spins on their Ising axis. Selecting
the [111] axis, in these two realizations a fourth of the spins are oriented
either parallel (AIAO) or antiparallel (AOAI) to the [111] axis.

\subsection{Energetics of the bulk phases in a field}
Whereas for Ising spins both realizations of the \AIAO order are equivalent,
which also remains true generally in absence of a magnetic field, this is not
the case in finite field. An effective canting of the spins out of their local
anisotropy axis, as a result of the dipolar-octupolar nature of the Nd$^{3+}$
ions, leads to a gain of Zeeman energy which can differ between the
configurations depending on the orientation of the magnetic field. Specifically,
the states split for $H\|$[111], but remain degenerate for $H\|$[001]
\cite{Opherden_2017-1}.

This can be seen considering the general model for Nd$^{3+}$ ions on the
pyrochlore lattice \cite{Huang_2014}
\begin{equation}
  H = \sum_{ij} J_{z} S^z_i S^z_j + J_{x} S^x_i S^x_j +J_{xz}S^x_i S^z_j + \sum_i g_{zz} \vect{B} \cdot \vect{e_i^z}  S^z_i
  \label{eq:ham}
\end{equation}
written with respect to the local basis. The $J_{xz}$ term can be removed by
tilting the spins in the local $x-z$ basis, thus, in the ground state
configuration the spins do not point along the local z-axis anymore. In the new
basis the model is characterised by the new couplings $\tilde{J_x}$,
$\tilde{J_z}$ and the angle $2\theta=\arctan (\frac{2 J_{xz}}{J_x-J_z})$.

We treat this microscopic hamiltonian classically and minimise the energy
obtaining the magnetic field dependence of the energy and magnetisation of the
AIAO and AOAI states.

For the related compound $\mathrm{Nd}_2\mathrm{Zr}_2\mathrm{O}_7$ this model has
been shown to give quantitative agreement with the experimental neutron
scattering results \cite{Benton_2016-1}. Since we do not have neutron scattering
data we are not able to directly extract the coupling parameters. Instead we
take the parameters obtained for $\mathrm{Nd}_2\mathrm{Zr}_2\mathrm{O}_7$
\cite{Benton_2016-1} as a starting point, noticing that there is some debate about the exact parameters \cite{Benton_2016-1, Petit_2016-2}. 

For \NHO we assume an effective moment of the Nd$^{3+}$ spins of
$\mu_{\mathrm{eff}} = 2.32 \mu_B$ related to $g_{zz}=2 \mu_{\mathrm{eff}}$. To
obtain an estimate of the other couplings, we compare the resulting
magnetisation curves with the experimental data for H$\||$[100]. Reasonable
agreement is achieved for $J_x=2.6 \times 0.103\, \mathrm{meV}$, $\tilde{J_y}=0
\, \mathrm{meV}$, $\tilde{J_z}= -2.6 \times 0.047\, \mathrm{meV}$, i.e. we
rescale the couplings obtained for $\mathrm{Nd}_2\mathrm{Zr}_2\mathrm{O}_7$ of
Ref.~\cite{Benton_2016-1} by a factor of 2.6, and $\theta =\pi/4$.

\begin{figure}
  \begin{minipage}{0.99\columnwidth}
    \includegraphics[width=.99\columnwidth]{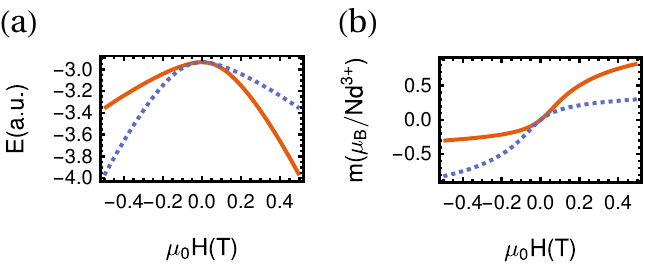}
  \end{minipage}
  \caption{Magnetic field dependence of the two realisations of the \AIAO order.  (a) Energy $E$ and (b) magnetisation $m$ of the AIAO (solid lines)
    and AOAI (dashed lines) state for the magnetic field $H$ oriented along [111].
    \label{fig:theory_E_and_M_AIAO}}
\end{figure}

Using these parameters we then consider the magnetic field dependence for fields
oriented along [111] where the experimental data shows hysteresis. In
Fig.~\ref{fig:theory_E_and_M_AIAO} we clearly observe an energy splitting (panel
a) of the two realizations of the \AIAO order and a different polarisability
(panel b) due to spin canting of the AIAO and AOAI state.
\subsection{Structure of spherical domain walls}
\begin{figure}[htbp]
	\begin{minipage}{0.22\textwidth}
		\includegraphics[width=\textwidth]{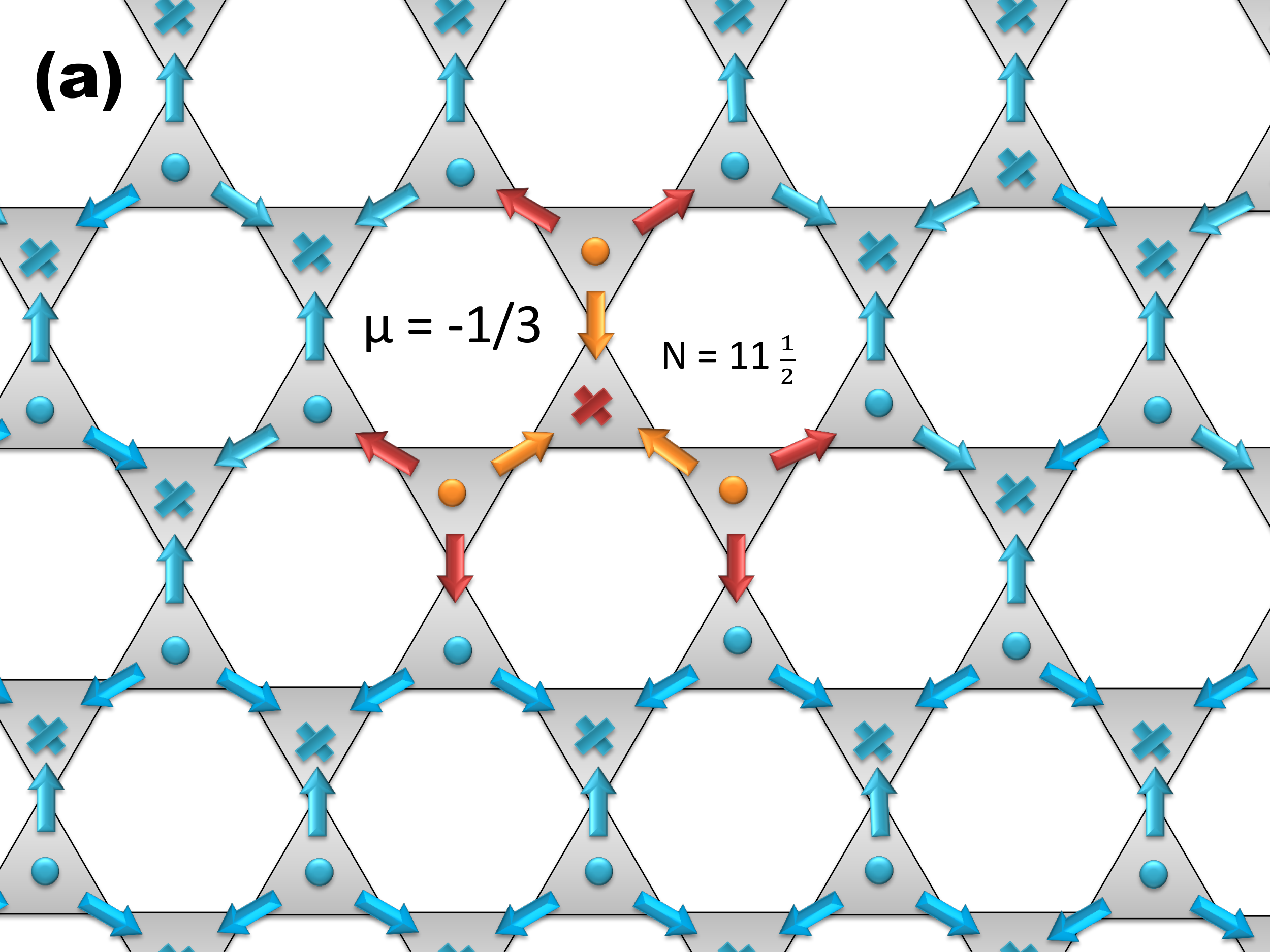}
	\end{minipage}
	\begin{minipage}{0.22\textwidth}
		\includegraphics[width=\textwidth]{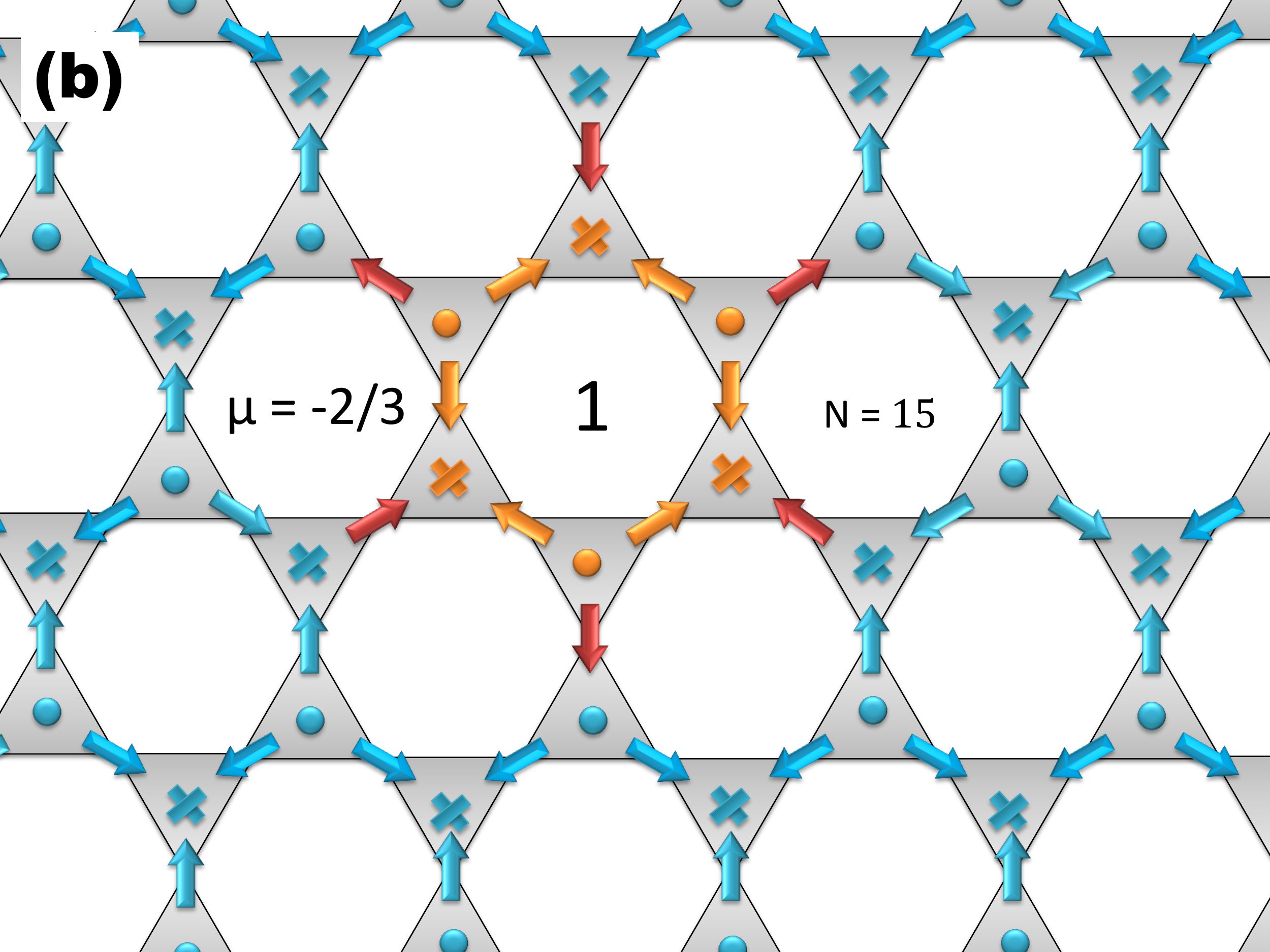}
	\end{minipage}
	\begin{minipage}{\textwidth}
		\hfill
	\end{minipage}
	\begin{minipage}{0.22\textwidth}
		\includegraphics[width=\textwidth]{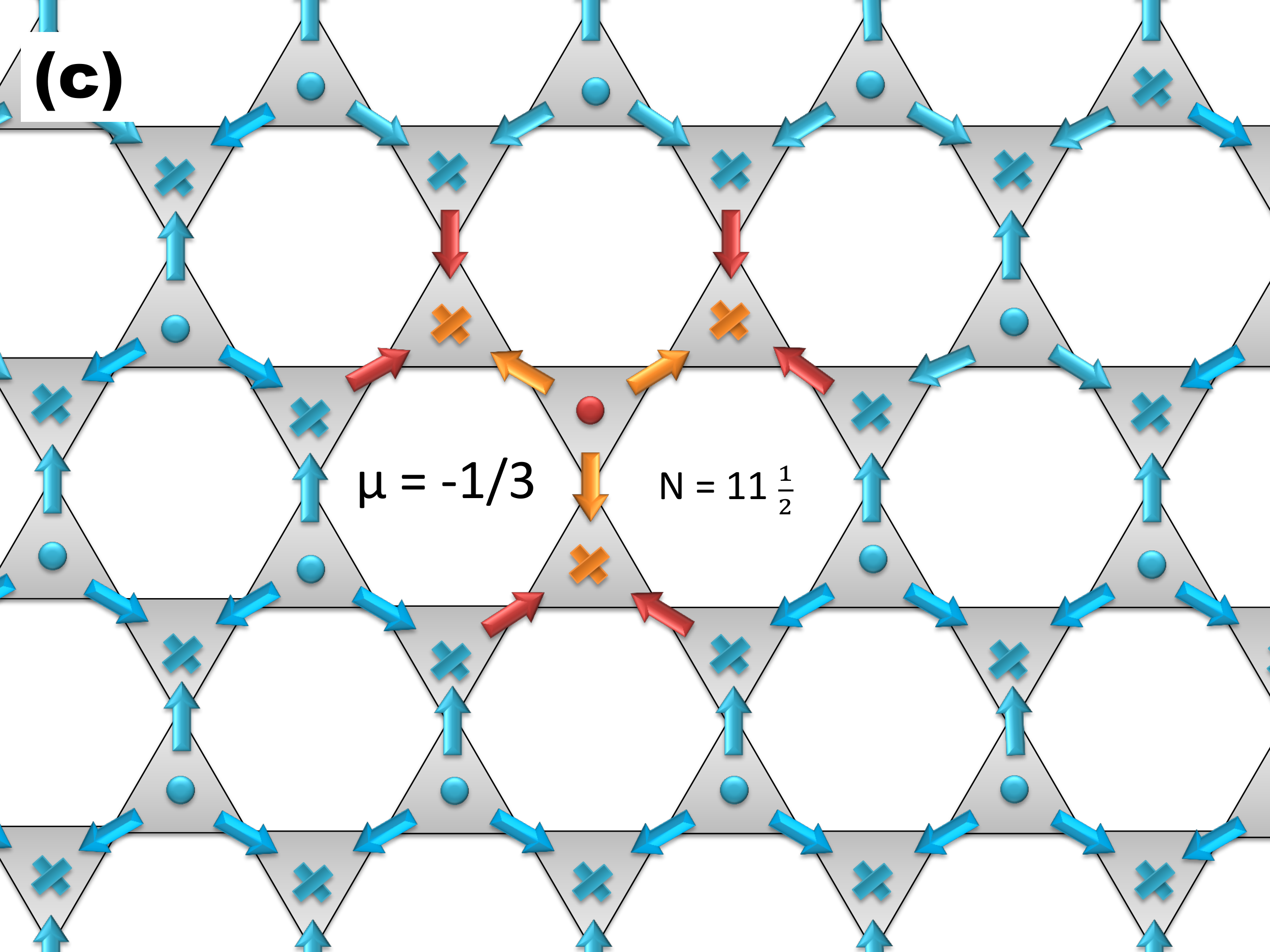}
	\end{minipage}
	\begin{minipage}{0.22\textwidth}
		\includegraphics[width=\textwidth]{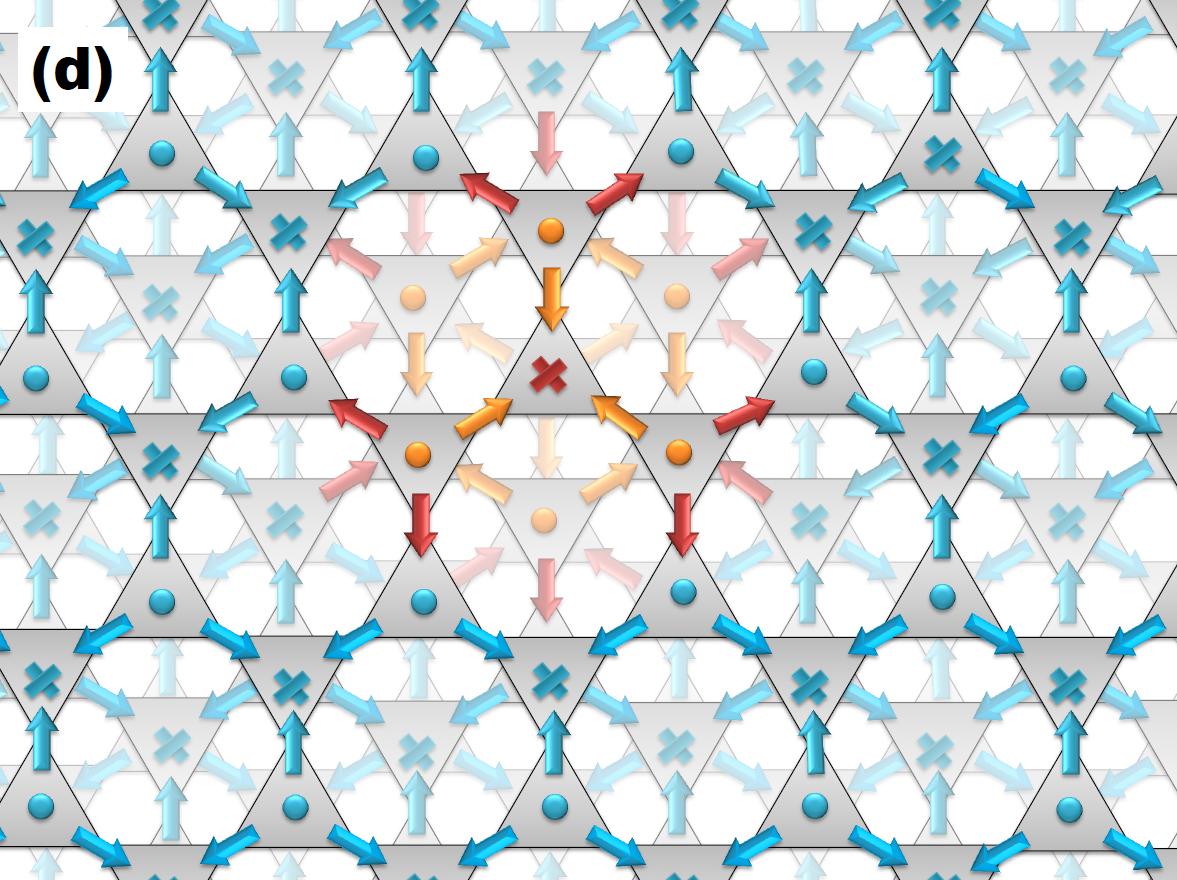}
	\end{minipage}
	\caption{Spherical domain with 3-In-1-Out boundaries oriented along [111],
    resulting in a negative magnetization in [111] direction. View onto the
    kagome layer of the pyrochlore lattice perpendicular to [1$\bar{1}$0]. (x -
    spin pointing in, $\bullet$ - spin pointing out). (a-c) view of different
    layers along [111], (d) full view}
	\label{fig:AIAO-Domaenenwaende}
\end{figure}

\begin{figure}
  \begin{minipage}{0.99\columnwidth}
    \includegraphics[width=.99\columnwidth]{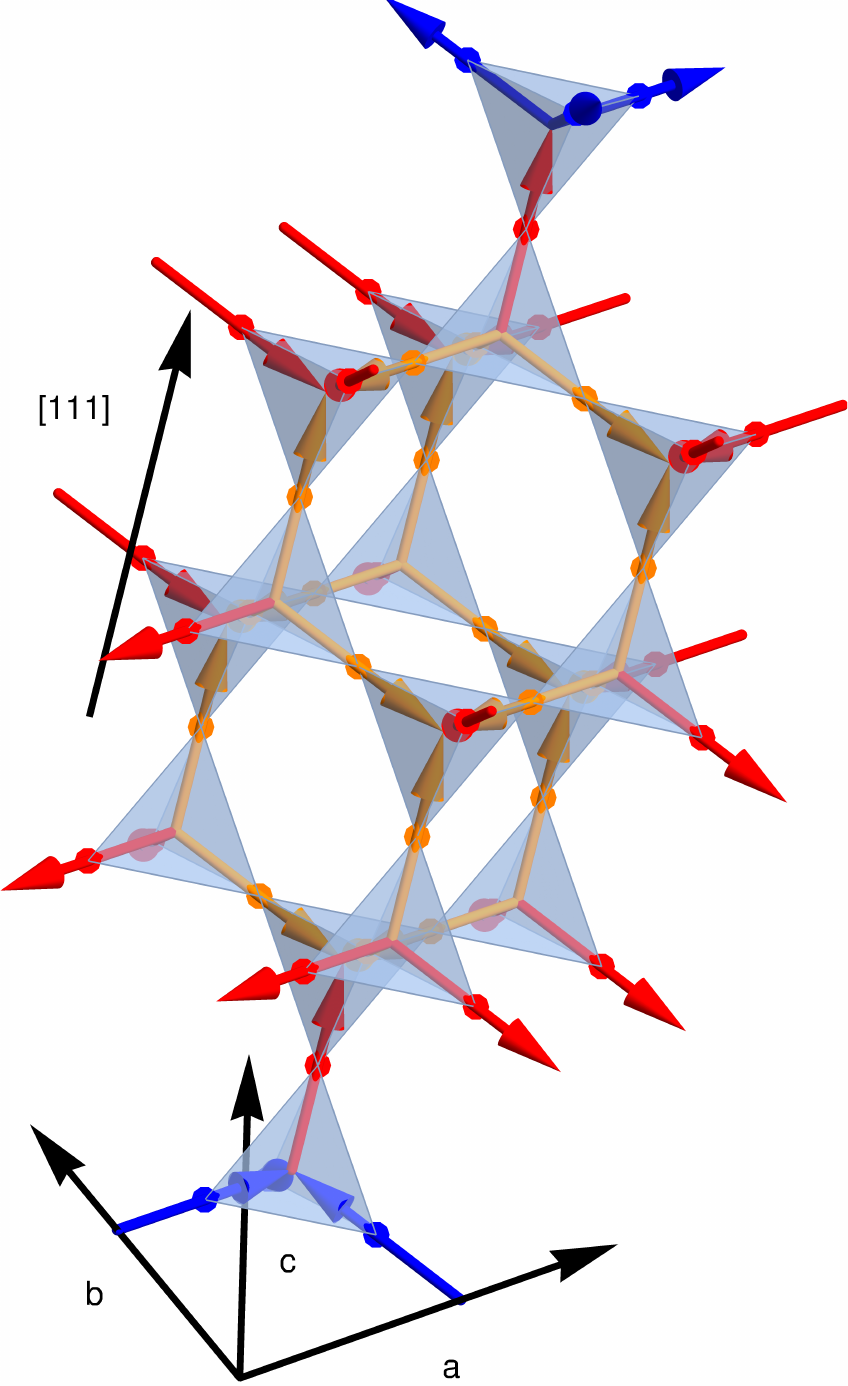}
  \end{minipage}
  \caption{3D view of spherical domain with 3-In-1-Out boundaries oriented along [111],
    resulting in a negative magnetization in [111] direction. Blue (orange) spins belong to AOAI (AIAO) bulk tetrahedra, red spins
		are boundary spins of the AIAO domain.
    \label{fig:supp_spherical_domain_3D}}
\end{figure}

The smallest closed spherical domain consists of 14 tetrahedra of AIAO (AOAI) in
a background of AOAI (AIAO). These are oriented along one of the equivalent
$\langle 111 \rangle$ directions, with a ring of 6 tetrahedra shrinking to 3 and
then 1 tetrahedron moving along the symmetry axis as depicted in
Fig.~\ref{fig:AIAO-Domaenenwaende} and Fig.~\ref{fig:supp_spherical_domain_3D}
for a domain with symmetry axis [111].
The panels in Fig.~\ref{fig:AIAO-Domaenenwaende} show the corresponding view of
the Kagome-layers of the pyrochlore lattice perpendicular to [1$\bar{1}$0]. The
symmetry equivalent spherical domains are constructed the same way by orienting
them along one of the four equivalent $\langle 111 \rangle$ axes.
Fig.~\ref{fig:supp_spherical_domain_3D} shows a 3D view of the same domain.

The depicted domain has 6 boundary tetrahedra with a moment of $2\,
\mu_{\mathrm{eff}}$ in [$\bar{1}$11], 6 with $2\,\mu_{\mathrm{eff}}$ in
[1$\bar{1}$1], 6 with $2 \mu_{\mathrm{eff}}$ in [11$\bar{1}$] and 2 with
$2\,\mu_{\mathrm{eff}}$ in [111] which results in a total magnetic moment of $-4
\,\mu_{\mathrm{eff}}$ in [111], and equivalently for the other symmetry related
domains. We emphasise that the total magnetic moment of an equal number of all
possible domains vanishes.

Using the microscopic hamiltonian (Eq.~\ref{eq:ham}) with the parameters as
discussed above we can estimate the energy cost and magnetisation per
tetrahedron of these domains as $E_{\mathrm{DW}} \approx 2 \, \mathrm{meV}$ and
$m_{\mathrm{DW}} \approx 0.3 \, \mu_{B}$.

\subsection{Energy barriers to domain wall growth}
Whereas in zero field all domains and the corresponding intermediate stages of
their growth have the same energy, in a finite field oriented along one of the
symmetry axes, e.g. [111], this is no longer the case. Note that for a field
applied along [001], the magnetic moments of the different spherical domains
projected onto the field direction is the same, and no specific domain is
favored.

To illustrate that the formation of negatively polarised spherical domains is
favoured in positive field we consider the specific case of a magnetic field
oriented along [111] and the domain in Fig.~\ref{fig:AIAO-Domaenenwaende} and
Fig.~\ref{fig:supp_spherical_domain_3D}. Considering the growth process along
the [111] axis we have the following stages: A single domain tetrahedra with a
vanishing moment, 4 flipped domain tetrahedra with a total moment of $2\,
\mu_{\mathrm{eff}}$, 10 domain tetrahedra with a moment of $0\,
\mu_{\mathrm{eff}}$, 13 domain tetrahedra with a moment of $-6\,
\mu_{\mathrm{eff}}$ and finally the close spherical domain with 14 domain
tetrahedra and a moment of $-4\, \mu_{\mathrm{eff}}$. We surprisingly observe
that the first stages of domain growth have the opposite magnetisation compared
to the final closed domain. The same is true for the alternative growth path,
starting with the ring of 6 tetrahedra with a moment of $4 \,
\mu_{\mathrm{eff}}$, adding each 3 tetrahedra above and below for a moment of
$-8 \, \mu_{\mathrm{eff}}$ and ending with the closed spherical domain with $-4
\, \mu_{\mathrm{eff}}$.

Thus, in a positive field along [111] the formation of the intermediate stages
of this type of domain is favoured compared to the ones oriented along the other
symmetry axes.

\subsection{Free energy}
Based on the observations made above, we propose a simple model free energy to
capture the inverted hysteresis. As the main non-equilibrium ingredient we
assume that in a preformed fully polarised phase of either AIAO or AOAI the
domain walls form predominantly with a magnetization opposite to the magnetic
field, due to energy barriers discussed above.

The free energy is constructed in the following way: The existence of domains is
due to the gain in entropy compared to a pure single domain state, and as the
most basic approximation we simply take the random mixing entropy
\begin{equation}
  S = x \ln(x) + (1-x) \ln(1-x) \, ,
\end{equation}
where $x$ is the volume fraction of one of the bulk phases.

Since in a finite field the AIAO/AOAI domains have different energies, we also
include an energy term for the bulk phases
\begin{equation}
  \begin{split}
    E_{\mathrm{Bulk}} &= x E_{\mathrm{AIAO}}(H) +(1-x) E_{\mathrm{AOAI}}(H)   \\
    &= x \left[ E_{\mathrm{AIAO}}(H)- E_{\mathrm{AOAI}}(H)\right] +
    \mathrm{const.}
  \end{split}
\end{equation}
where we obtain the different field dependent energies for the two realisations
of the \AIAO order from a classical treatment of the microscopic Hamiltonian
\ref{eq:ham}.

In addition, the domain walls between the bulk phases contribute to the energy.
The surface in random mixing is proportional to $x(1-x)$ which results in an
energy cost
\begin{equation}
  E_{\mathrm{S}}= x(1-x)\left[E_{\mathrm{DW}}- m_{\mathrm{DW}} H \right] \, ,
\end{equation}
with a constant contribution $E_{\mathrm{DW}}$ due to broken bonds at the domain
boundary and a Zeeman energy due to the domain wall magnetization
$m_{\mathrm{DW}}$.

The domain wall surface magnetization, $m_{\mathrm{DW}}$, is the crucial
non-equilibrium input, we take it to be opposite to the magnetic field direction
due to the energy barriers to domain wall growth which favor this type of
domains. Moreover, it depends on the initially prepared state: it takes its
maximal value for the fully polarised initial condition, and decreases down to
zero for the unpolarised condition.

In total we model the free energy for a mixture of AIAO and AOAI domains with
domain walls as
\begin{align}
  f &=  x \left[ E_{\mathrm{AIAO}}(H)- E_{\mathrm{AOAI}}(H)\right]+x(1-x) \left[E_{\mathrm{DW}}- m_{\mathrm{DW}} H \right] \notag\\
    &+T \left[ x \ln(x) +(1-x) \ln(1-x)\right]
      \label{eq:supp_free_energy}
\end{align}
\begin{figure}
  \begin{minipage}{0.99\columnwidth}
    \includegraphics[width=.99\columnwidth]{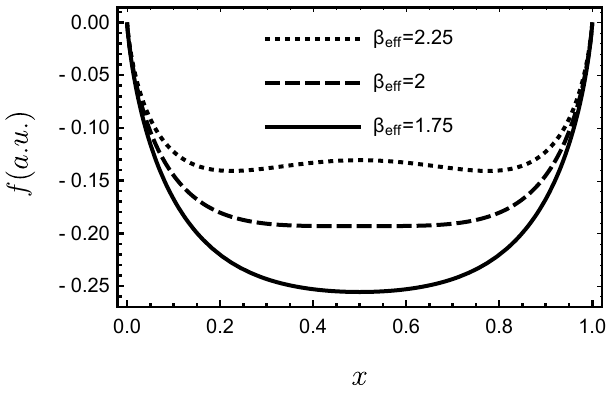}
  \end{minipage}
  \caption{Phase transition occurring in the model free energy. The free energy
    $F$ shows qualitatively different behavior as a function of the bulk phase
    volume fractions $x$ when changing the parameters $\beta_{\mathrm{eff}}=
    (\beta+m_{\mathrm{DW}} H)/T$, here shown for $\alpha=0$. Below
    $\beta_{\mathrm{eff}}=2$ we obtain a unique equal-density mixture phase, whereas above
    there are two high-low density mixture phases.
    \label{fig:free_energy_transition}}
\end{figure}
Let's first discuss the case of balanced bulk phases, e.g. no spin canting,
hence, no energy difference between the bulk states
($E_{\mathrm{AIAO}}=E_{\mathrm{AOAI}}$). In that case we obtain a
phase-transition as a function of $\beta_{\mathrm{eff}}=(E_{\mathrm{DW}} -
m_{\mathrm{DW}}H)/T$ between a unique equal mixture with $x=0.5$ for
$\beta_{\mathrm{eff}} <2 $ , and a high/low density phase with $x \ne 0.5$ and
two solutions for $\beta_{\mathrm{eff}} >2$ as shown in
Fig.~\ref{fig:free_energy_transition}.

Considering imbalanced bulk states, e.g. assuming a cubic energy difference
($E_{\mathrm{AIAO}}-E_{\mathrm{AOAI}}=\alpha H^3$), one of the solutions becomes
meta-stable up to a critical field strength where a transition into a unique
high/low density-phase occurs.
\begin{figure}
  \includegraphics[width=.99\columnwidth]{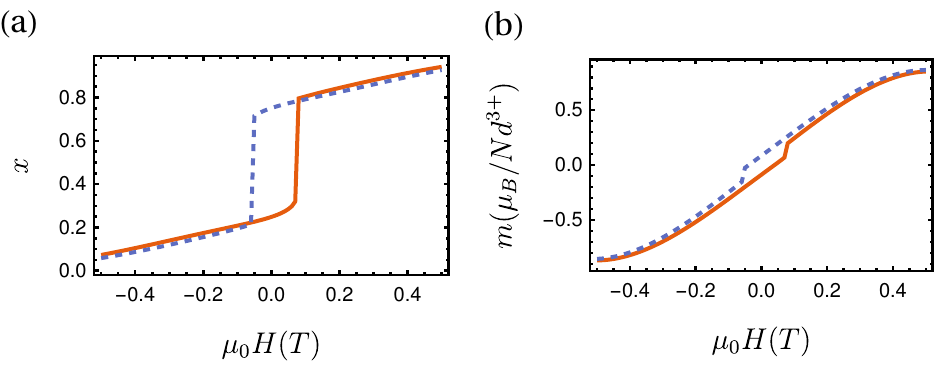}
  \caption{Magnetic field dependence resulting from the model free energy for negatively
    (dashed) and positively (solid) polarised initial states ($m_{\mathrm{DW}}=
    \pm 2.1$). (a) Bulk phase volume fractions $x$ and (b) magnetization $m$
    (including the bulk and domain wall contributions) as a function of magnetic field $H$.
    \label{fig:toy_hysteresis}}
\end{figure}

This simple model naturally leads to an inverted hysteresis loop,
Fig.~\ref{fig:toy_hysteresis}. For the positively polarised initial condition
($m_{\mathrm{DW}}< 0$, solid curve in Fig.~\ref{fig:toy_hysteresis}(a)) the
volume fraction evolves smoothly from $x=1$ at large positive field down to a
critical negative field where the phase becomes unstable and the volume fraction
jumps, and then connects smoothly to the $x=0$ negatively polarised phase.
Similarly, for the negatively polarised initial condition ($m_{\mathrm{DW}} >0$,
dashed curve in Fig.~\ref{fig:toy_hysteresis}(a)) the state evolves smoothly
from $x=0$ at large negative fields up to a critical positive field where the
volume fraction jumps, and then connects to the positively polarised state. Due
to the domain wall magnetization the resulting magnetization curve shows an
inverted hysteresis (Fig.~\ref{fig:toy_hysteresis}(b)) with a remanence at $H=0$
which is negative (positive) for the positively (negatively) polarised initial
states (solid/dashed curves respectively). We emphasise that this toy model
demonstrates that the basic phenomenology described here does not depend on the
specific choices made for the comparison with the experimental data detailed
next.

Finally, we turn to the comparison with the experiment. We obtain the energy
(and magnetisation dependence) of the bulk phases
$E_{\mathrm{AIAO}/\mathrm{AOAI}}(H)$ from a classical treatment of the
microscopic Hamiltonian as described above. Based on the discussed spherical
domain structure and the microscopic hamiltonian we obtain $E_{\mathrm{DW}}= 2.3
\, \mathrm{meV}$ and $m_{\mathrm{DW}}=\pm 0.3 \mu_{B}$ (see above). The absolute
temperature, and hence the absolute energy scale, remains a free parameter since
the entropy of the resulting mixture is only roughly approximated by the random
mixing entropy and would depend on the specific configuration, sizes and forms
of the domains.

\begin{figure}
  \includegraphics[width=.99\columnwidth]{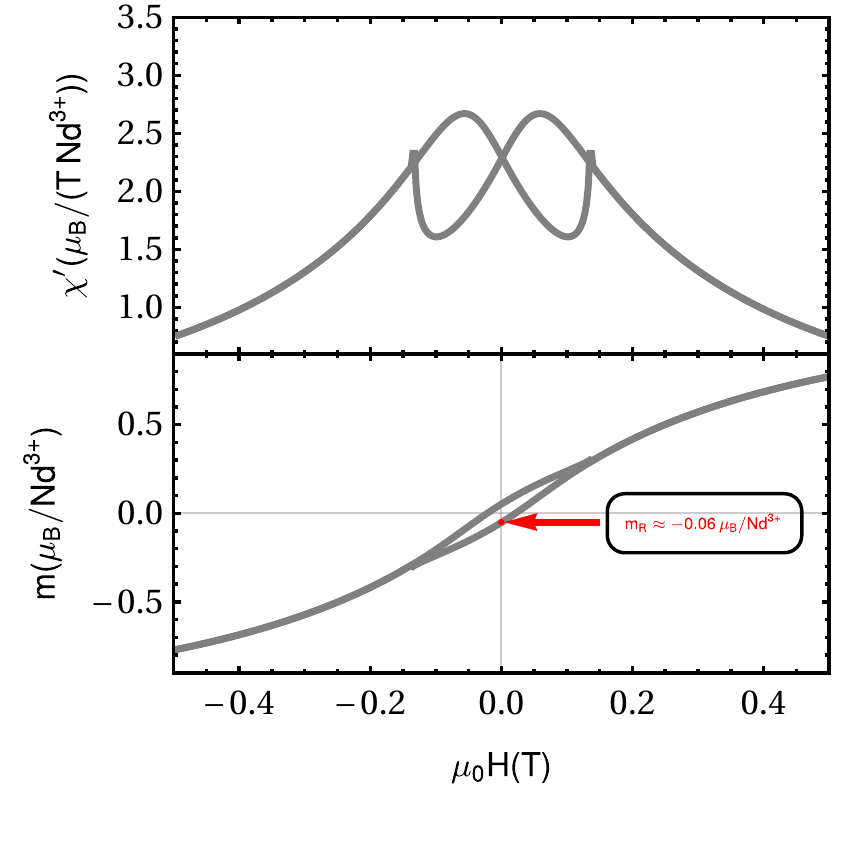}
  \caption{Theoretical magnetic-field dependence of (a) the susceptibility 
    and (b) the magnetisation.
    Results obtained from the proposed model free energy using the energy and
    magnetisation curves of the AIAO and AOAI states obtained from the classical treatment
    of the microscopic hamiltonian.
    \label{fig:theory_hysteresis}}
\end{figure}

Using above parameters in the model free energy (Eq.~\ref{eq:supp_free_energy})
we obtain the hysteresis behavior shown in Fig.~\ref{fig:theory_hysteresis} and
in the main text.

\end{document}